\documentclass[prb,twocolumn,amsmath,amssymb,superscriptaddress]{revtex4-2}
\usepackage{epsfig,graphicx,graphics,float}
\usepackage[T1]{fontenc}
\usepackage[utf8]{inputenc}
\usepackage{appendix}
\usepackage{amscd}
\usepackage{bm}
\usepackage{psfrag} 
\usepackage{bbm} 
\usepackage{babel}
\usepackage{wasysym}
\usepackage{mathrsfs}
\usepackage{xcolor}
\usepackage{array}
\usepackage{subfigure}
\usepackage{verbatim} 
\usepackage{braket}
\usepackage{soul}
\usepackage[normalem]{ulem}
\usepackage[final]{hyperref}

\hypersetup{
	colorlinks=true,       
	linkcolor=blue,       
	citecolor=blue,
	filecolor=magenta,    
	urlcolor=blue         
}

\newcommand{\be}{\begin{equation}}
\newcommand{\ee}{\end{equation}}
\newcommand{\bea}{\begin{eqnarray}}
\newcommand{\eea}{\end{eqnarray}}
\newcommand{\bmat}{\begin{pmatrix}}
\newcommand{\emat}{\end{pmatrix}}
\newcommand{\lb}{\left(}
\newcommand{\rb}{\right)}
\newcommand{\lsb}{\left[}
\newcommand{\rsb}{\right]}
\newcommand{\mc}{\mathcal}
\newcommand{\mb}{\mathbf} 
\newcommand{\mr}{\mathrm} 
\newcommand{\trup}{\bigtriangleup}
\newcommand{\trdo}{\bigtriangledown}
\DeclareMathOperator{\real}{\mr{Re}}
\DeclareMathOperator{\imag}{\mr{Im}}

\newcommand{\new}[1]{\textcolor{black}{#1}}

\begin{document}
\title{Compact localized fermions and Ising anyons in a chiral spin liquid}

\author{Tim Bauer}
\affiliation{Dahlem Center for Complex Quantum Systems and Fachbereich Physik, Freie Universit\"at Berlin, Arnimallee 14, 14195 Berlin, Germany}
\affiliation{Helmholtz-Zentrum Berlin f\"ur Materialien und Energie, Hahn-Meitner-Platz 1, 14109 Berlin, Germany}
\affiliation{Department of Applied Physics, The University of Tokyo, Bunkyo, Tokyo 113-8656, Japan}
\thanks{JSPS Postdoctoral Fellow for Research in Japan}

\author{Johannes Reuther}
\affiliation{Dahlem Center for Complex Quantum Systems and Fachbereich Physik, Freie Universit\"at Berlin, Arnimallee 14, 14195 Berlin, Germany}
\affiliation{Helmholtz-Zentrum Berlin f\"ur Materialien und Energie, Hahn-Meitner-Platz 1, 14109 Berlin, Germany}

\begin{abstract}
    Quasiparticle hybridization remains a major challenge to realizing and controlling exotic states of matter in existing quantum simulation platforms.
    We report the absence of hybridization for compact localized states (CLS) emerging in the chiral spin liquid described by the Yao-Kivelson model.
    The CLS form due to destructive quantum interference at fine-tuned coupling constants and populate perfectly flat quasiparticle bands on an effective kagome lattice.
    Using a formalism for general Majorana-hopping Hamiltonians, we derive exact expressions for CLS for various flux configurations and both for the topological and trivial phases of the model.
    In addition to finite-energy matter fermions with characteristic spin-spin correlations, we construct compact localized Majorana zero modes attached to $\pi$-flux excitations, which enable non-Abelian braiding of Ising anyons with minimal separation.
    Our results inform the quantum simulation of topologically ordered states of matter and open avenues for exploring flat-band physics in quantum spin liquids.
\end{abstract}
\maketitle

\section{Introduction}
    \label{sec:1}
    The remarkable progress in the quantum simulation and preparation of topologically ordered states of matter opens new avenues for experimental tests of quantum spin liquids (QSLs) \cite{semeghini2021probing,satzinger2021realizing,kalinowski2023non,will2025probing,evered2025probing}.
    In particular, recent claims of successful braiding of non-Abelian anyons in quantum simulators \cite{google2023non,xu2023digital,xu2024non,iqbal2024non,iqbal2025qutrit,song2025shortcuts,evered2025probing} hold promise for the control of fractional excitations hosted by QSLs and constitute a milestone on the long path to robust quantum information processing. 
    These braiding experiments were based on the stabilization of toric-code or string-net states in superconducting \cite{google2023non,xu2023digital,xu2024non} or trapped-ion \cite{iqbal2024non,iqbal2025qutrit} quantum processors comprising between $20$ and $70$ qubits or qudits. 
    With the rich zoo of theoretically proposed QSLs in mind \cite{savary2016quantum,broholm2020quantum}, this raises the question of which spin model can  be simulated at these intermediate scales without significant hybridization of its fractional excitations. 
    As the hybridization of particles is typically governed by their capability to localize, these considerations naturally point to the study of the limiting case of \emph{compact localization}.

    Compact localized states (CLS) are eigenmodes with support on a finite number of lattice sites, and stem from {destructive} quantum interference on specific lattice geometries \cite{kirkpatrick1972localized,sutherland1986localization,zhang2020compact,bergman2008band}.
    {If this destructive interference is realized through fine-tuning of an applied magnetic flux, the appearance of CLS is known as Aharonov–Bohm caging \cite{vidal1998aharonov}, and has been experimentally confirmed in photonic and cold-atom platforms \cite{mukherjee2018experimental,li2022aharonov,chen2025interaction}.}
    {Moreover, since CLS} do not disperse, they form perfectly flat energy bands in crystallized systems, and thereby give rise to multiple phenomena of persistent interest to the condensed matter community.
    Electronic systems with low-energy flat bands are susceptible to instabilities which induce unconventional phases and, in some instances, fractionalization of microscopic degrees of freedom.
    Currently prominent examples of this scenario are given by the unconventional superconductivity in kagome metals \cite{yin2022topological,wang2023quantum,wang2024topological} and the fractional Chern insulating phase reported in Moire systems \cite{xie2021fractional,bhowmik2024emergent}.
    On the other hand, in the context of strongly correlated materials, magnonic flat bands play a crucial role in the explanation of antiferromagnetism on frustrated lattices \cite{schulenburg2002macroscopic,schmidt2006linear,bergman2008band}.
    Moreover, theoretical results indicate that the response of perfectly flat bands to disorder should exhibit surprising localization effects, such as an \emph{inverse} Anderson transition in three dimensions \cite{goda2006inverse} and critical states in two dimensions \cite{chalker2010anderson}.

    Motivated by this rich variety of flat-band physics and by the relevance for quantum simulations, we study flat bands formed by compact localized fractional excitations in QSLs.
    To obtain exact analytical expressions, we employ the exactly solvable Kitaev model adapted to the star lattice \cite{kitaev2006anyons,yao2007exact}.
    Originally defined on the honeycomb lattice, the Kitaev model provides a powerful theoretical benchmark of $\mathbbm Z_2$ QSLs as well as of non-Abelian Ising order \cite{kitaev2006anyons}.
    Moreover, the theoretical prediction \cite{khaliullin2005orbital,jackeli2009mott} of the characteristic exchange anisotropy of the model in spin-orbit-assisted Mott insulators initiated enormous experimental and theoretical efforts to realize a QSL phase in various transition metal compounds,  for recent reviews see Refs.~\cite{winter2017models,hermanns2018physics,trebst2022kitaev}.
    While the most-studied candidate materials, such as $\alpha$-RuCl$_3$, crystallize in two-dimensional honeycomb layers, one also expects the characteristic spin exchange and the resulting spin-liquid physics in compounds with other two- or three-dimensional lattice structures \cite{trebst2022kitaev}.
    This is because the Kitaev model can be defined and solved on any tricoordinated graph.
    In fact, the first example of this is given by the adaptation of the model on the star lattice, commonly referred to as the Yao-Kivelson model \cite{yao2007exact}.
    An important difference between the Kitaev honeycomb model and the Yao-Kivelson model is that the ground state of the latter spontaneously breaks time-reversal symmetry and is, therefore, identified as a chiral spin liquid.
    This remarkable result can be attributed to the non-bipartite geometry of the star lattice \cite{kitaev2006anyons,yao2007exact}.
    
    In this work, we point out that the same geometry also allows for the destructive quantum interference necessary for the formation of CLS.
    Accordingly, we find that upon fine tuning the coupling ratio, the energy dispersion of fermionic excitations of the Yao-Kivelson model vanishes identically.
    The resulting perfectly flat bands are found both in the topological and trivial phase of the model and for various uniform flux patterns, including the $\pi$-flux-free ground-state sector.
    While finite-energy flat bands are formed by compact localized matter fermions with characteristic spin-spin correlations, we also find a flat band populated by compact localized Majorana zero modes (MZMs) on plaquettes pierced by a $\pi$ flux.
    MZMs attached to low-energy $\pi$ flux excitations on top of the ground state therefore form \emph{compact localized Ising anyons} at the specific value for the coupling ratio.
    The compact nature of these anyons implies the absence of any hybridization of MZMs and allows for non-Abelian braiding of anyons with minimal separation.
    While recent works studied magnonic flat bands in the field-polarized Yao-Kivelson model \cite{zou2025high}, or flat bands of unpaired Majorana modes of the periodically depleted Kitaev honeycomb model \cite{fukui2025topological}, this work is, to the best of our knowledge, the first detailed study of {perfectly} flat bands and compact localization of itinerant fractional excitations in QSLs.
    Our work can be readily adapted to other lattice geometries or more general Majorana hopping models, and it lays the groundwork for the exploration of flat-band phenomenology in QSLs.
    {We note that a recent related work studied the construction of Majorana Wannier orbitals in staggered flux patterns for the Kitaev honeycomb model and also reported quasiparticle bands with suppressed width \cite{yogendra2025fractional}.}
    
    The remainder of this paper is structured as follows. In Sec.~\ref{sec:2}, we introduce the Yao-Kivelson model, outline its exact solution and discuss the energy bands of various uniform flux sectors, focusing on perfectly flat bands.
    Subsequently, we derive the conditions for CLS in generic Majorana hopping models and detail the wavefunction of the CLS at the coupling ratios of the identified flat bands in Sec.~\ref{sec:3}.
    We compute the characteristic equal-time spin-spin correlations of CLS in Sec.~\ref{sec:4}, and discuss our findings and interesting research directions in Sec.~\ref{sec:5}.
    The Appendices~\ref{sec:a} and \ref{sec:b} provide more details on the exact solution of the model and the eigenproblem for CLS, respectively.
    
\section{Model and fermionic spectrum}
    \label{sec:2}
\subsection{Majorana representation}
    \begin{figure}
        \centering
        \includegraphics[width=\columnwidth]{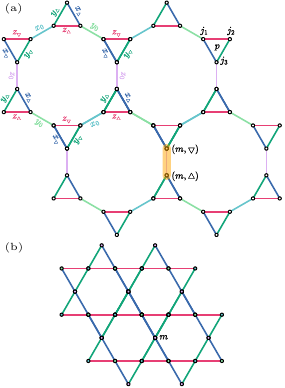}
        \caption{(a) Star lattice with nine bond types $\alpha_\chi$. The lattice sites $j_1$, $j_2$ and $j_3$ wind around the triangular plaquette $p$ following the convention in Eq.~\eqref{eq:wilson}. Itinerant Majoranas on the orange-shaded sites $(m, \trdo)$ and $(m, \trup)$ are paired in Eq.~\eqref{eq:matter_pairing}. (b) The paired Majoranas form a fermion located on site $m$ of the kagome lattice.}
        \label{fig:1}
    \end{figure}
    The Yao-Kivelson model \cite{yao2007exact} is defined on the star (also known as triangle-honeycomb) lattice.
    As shown in Fig.~\ref{fig:1}(a), the unit cell of this lattice comprises nine different bond types which we label by $x_\trdo$, $y_\trdo$, $z_\trdo$ and $x_\trup$, $y_\trup$, $z_\trup$
    and $x_0$, $y_0$, $z_0$.
    The model Hamiltonian is given by
    \be
        \label{eq:model}
        H=J_\trdo\sum_{\braket{jk}^{\alpha}_\trdo}\sigma^\alpha_j\sigma^\alpha_k+J_\trup\sum_{\braket{jk}_{\trup}^\alpha}\sigma^\alpha_j\sigma^\alpha_k+J_0\sum_{\braket{jk}_{0}^\alpha}\sigma^\alpha_j\sigma^\alpha_k,
    \ee
    where $\sigma^\alpha_j$, with $\alpha\in\{x,y,z\}$, is a Pauli matrix acting on a spin-$1/2$ degree of freedom on site $j$, and $\braket{jk}^\alpha_\chi$, with $\chi\in\{\trdo,\trup,0\}$, denotes the nearest-neighbor bond of type $\alpha_\chi$ connecting the sites $j$ and $k$.
    {The summation is understood to include both $\braket{jk}^\alpha_\chi$ and $\braket{kj}^\alpha_\chi$, i.e., double counting of all bilinears $\sigma^\alpha_j\sigma^\alpha_k$.
    In this work, we focus on isotropic intra-triangle coupling $J_\trdo=J_\trup$, following Ref.~\cite{yao2007exact}.}
    
    Since the star lattice is tricoordinated, we can obtain the full exact spectrum of $H$ in terms of Kitaev's paradigmatic solution for the equivalent model on the honeycomb lattice \cite{kitaev2006anyons}.
    This approach is based on the representation of the spin degree of freedom
    \be
        \label{eq:spin_majorana}
        \sigma^\alpha_j=ic_j^\alpha c_j,
    \ee
    where $c_j^x$, $c_j^y$, $c_j^z$ and $c_j$ are Majorana fermion operators subject to the local constraint $c_j^xc_j^yc_j^zc_j=1$.
    This constraint is necessary to restrict solutions to the physical Hilbert space and  controls the parity of fermionic quasiparticle excitations \cite{kitaev2006anyons,pedrocchi2011physical,zschocke2015physical}.
    
    Within the Majorana representation~\eqref{eq:spin_majorana}, we introduce the bond variables $U_{jk}=ic_j^\alpha c_k^\alpha$ for every bond $\braket{jk}^\alpha_\chi$ to describe an emerging static $\mathbbm Z_2$ gauge field.
    The static nature of the gauge field follows from the vanishing commutators of bond variables with other bond variables and the Hamiltonian~\eqref{eq:model}, and is associated with the conserved flux through the elementary plaquettes of the lattice.
    As illustrated in Fig.~\ref{fig:1}(a), the star lattice is composed both of triangular plaquettes with length $\ell=3$ and dodecagonal plaquettes with length $\ell=12$.
    For any given plaquette $p$, the conserved flux is described by the (gauge-invariant) plaquette operators
    \be
        \label{eq:wilson}
        W_p=(-i)^{\ell_p}\prod_{n=1}^{\ell_p} U_{j_n,j_{n+1}},
    \ee
    where $\ell_p$ is the length of $p$ and the ordered set $(j_1,...,j_{\ell_p})$ describes a loop winding around the boundary of $p$ clockwise using the convention $j_{\ell_p+1}=j_1$, see Fig.~\ref{fig:1}(a). 
    It is straightforward to show that plaquette operators on dodecagonal and triangular plaquettes have eigenvalues $\pm 1$ and $\pm i$, respectively.
    The latter is a consequence of {the odd loop length of triangles} and indicates a spontaneous breaking of time-reversal symmetry, inducing a chiral spin liquid \cite{kitaev2006anyons,yao2007exact}.
    Moreover, we note that using the convention in Eq.~\eqref{eq:wilson}, a plaquette operator with eigenvalue $+1$ corresponds to a $\pi$ flux.
    
    We can solve for the spectrum by choosing a gauge configuration of the flux sector of interest, i.e., replacing the bond variables $U_{jk}$ by their eigenvalues $u_{jk}=\pm 1$ (with $u_{kj}=-u_{jk}$) and thereby rendering the Hamiltonian~\eqref{eq:model} quadratic in the itinerant Majoranas $c_j$.
    More generally, we may study the Majorana hopping model
    \be
        \label{eq:majorana_hopping}
        H_{t}=i\sum_{j,k} t_{jk}c_jc_k,
    \ee
    parametrized by an antisymmetric hopping matrix $t$ with elements $t_{jk}=-t_{kj}$.
    In the case of the Yao-Kivelson model, finite matrix elements are given by $t_{jk}=-u_{jk}J_\chi$ (with $J_\trup=J_\trdo$) for nearest neighbors coupled by the bond $\braket{jk}_\chi^\alpha$.
    However, it is straightforward to adapt many of our results for different geometries or for more general hopping matrices.
    
    In the context of flat-band physics, it is instructive to rewrite the hopping model~\eqref{eq:majorana_hopping} in terms of complex \emph{matter} fermions by pairing neighboring Majoranas on different triangles.
    Labeling sites on down-pointing triangles by tuples $j=(m,\trdo)$ and their nearest neighbors on up-pointing triangles by $k=(m,\trup)$, we introduce the canonical annihilation and creation operators
    \be
        \label{eq:matter_pairing}
        f_m=\frac{1}{2}\lb c_{m,\trdo}+ic_{m,\trup}\rb,\quad f_m^\dagger=\frac{1}{2}\lb c_{m,\trdo}-ic_{m,\trup}\rb,
    \ee
    respectively.
    As indicated in Fig.~\ref{fig:1}(b), the prescribed pairing process results in an effective description on the kagome lattice.
    Assuming nearest-neighbor hopping $t$, we can rewrite Eq.~\eqref{eq:majorana_hopping} to
    \bea
        \label{eq:matter_hopping}
        H_t&=&\sum_{\braket{m,n}'} \lb\tau^+_{mn} f_m^\dagger f_n+\tau^-_{mn} f_m^\dagger f_n^\dagger+\mr{h.c.}\rb\nonumber\\
        &+&\sum_{m} \mu_m\lb f_m^\dagger f_m-\frac{1}{2}\rb,
    \eea
    where $\braket{m,n}'$ denotes nearest-neighbor bonds on the kagome lattice, $\tau_{mn}^\pm=i\lb t_{m,\trdo
    ,n,\trdo}\pm t_{m,\trup,n,\trup}\rb$ corresponds to normal and anomalous hopping, respectively, and $\mu_m=2t_{m,\trdo,m,\trup}$ is an effective on-site potential.
    We note that the emergence of flat bands in metallic, superconducting or magnonic states on the kagome lattice has been and remains a subject of intensive research \cite{schulenburg2002macroscopic,schmidt2006linear,bergman2008band,yin2022topological,wang2023quantum,wang2024topological}.
    In this context, Eq.~\eqref{eq:matter_hopping} provides an addition to the list of tight-binding models hosting perfectly flat bands, but with parameters encoding an underlying flux sector.

    As specified in Appendix~\ref{sec:a}, we obtain the spectrum of Eq.~\eqref{eq:matter_hopping} for arbitrary parameters using the Bogoliubov-de-Gennes formalism.
    In the remainder of this section, however, we focus on uniform flux patterns and translationally invariant hopping.
    
\subsection{Fractional flat bands}
    \begin{figure*}
        \centering
        \includegraphics[width=1\linewidth]{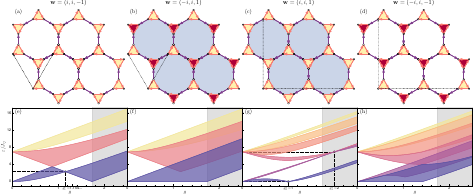}
        \caption{(a)-(d) Gauge configurations of uniform flux patterns (a) $\mb w=(i,i,-1)$, (b) $(-i,i,1)$, (c) $(i,i,1)$ and (d) $(-i,i,-1)$. A solid arrow pointing from site $j$ to site $k$ indicates $u_{jk}=+1=-u_{kj}$. White (blue) shaded dodecagons are pierced by zero ($\pi$) flux. Yellow (red) shaded triangles are pierced by $\pi/2$ ($-\pi/2$) flux. The black dashed arrows mark the primitive lattice vectors of the resulting Majorana hopping model~\eqref{eq:majorana_hopping}. (e)-(h) Single-particle energies $\varepsilon$ of corresponding quasiparticle bands as function of the coupling ratio $g=J_0/J_\trdo$. 
        {The gray (white) shaded background represents the trivial (topological) phase \new{of the global ground state in the sector $\mb w=(i,i,-1)$}.}
        Black circles indicate the flat bands specified in (e) Eq.~\eqref{eq:flat1} and (g) Eqs.~\eqref{eq:flat2} and \eqref{eq:flat3}. {The energy bands are symmetric with respect to $g\rightarrow -g$.}}
        \label{fig:2}
    \end{figure*}
    Since the star lattice is non-bipartite, Liebs theorem \cite{lieb1994flux,kitaev2006anyons} cannot be applied to determine the flux pattern of the ground state of the full model~\eqref{eq:model} and instead, one has to numerically compare the energies of different flux sectors \cite{yao2007exact} {or employ perturbation theory for either dominating intra- or inter-triangle coupling \cite{dusuel2008perturbative}}.
    Focusing on uniform flux patterns, we distinguish three types of elementary plaquettes - down-pointing triangles, up-pointing triangles and dodecagons - and equate the eigenvalues of all plaquette operators of the same type with $w_\trdo=\pm i$, $w_\trup=\pm i$ and $w_0=\pm 1$, respectively.
    By varying the triple $\mb w=(w_\trdo, w_\trup, w_0)$ and up to degenerate time-reversal partner states, we arrive at the four uniform flux sectors parametrized by $\mb w=(i,i,-1)$, $(-i,i,1)$, $(i,i,1)$ and $(-i,i,-1)$ in Figs.~\ref{fig:2}(a)-(d).
    For {periodic boundary conditions and} the gauge configurations indicated in the respective figures, the hopping models~\eqref{eq:majorana_hopping} of all four sectors exhibit translational invariance and can be diagonalized using a Fourier transformation. 

    The resulting energy bands of the fermionic quasiparticle excitations are shown in Figs.~\ref{fig:2}(e)-(h) as a function of the coupling ratio $g=J_0/J_\trdo$.
    We note that the hopping models for the flux patterns $\mb w=(\pm i,i,\pm 1)$ have an enlarged unit cell on a rectangular lattice, see Figs.~\ref{fig:2}(g), (h), and therefore host twice as many bands compared to the hopping models for the sectors $(\pm i,i,\mp 1)$ with elementary unit cells on a triangular lattice \cite{tikhonov2010quantum}.
    Moreover, we note that the sectors with different eigenvalues for down- and up-pointing triangles, i.e., with $w_\trdo=-w_\trup$, are odd under inversion symmetry.
    As a consequence, a valley degree of freedom associated with the two sublattices of the undressed honeycomb lattice is activated and allows for the formation of a neutral Fermi surface in Figs.~\ref{fig:2}(f), (h), see also Refs. \cite{tikhonov2010quantum,fuchs2020parity,chari2021magnetoelectric,nakazawa2022asymmetric}.

\begin{figure}
        \centering
        \includegraphics[width=\columnwidth]{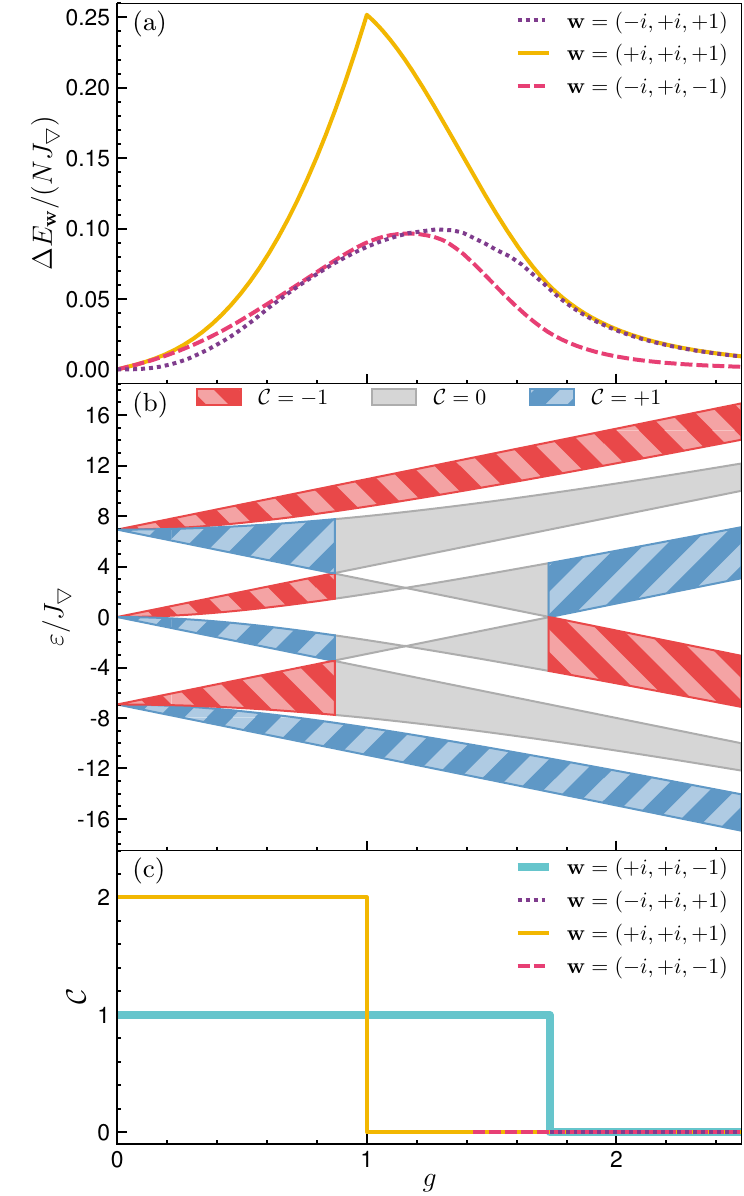}
        \caption{(a) Energy difference $\Delta E_{\mb w}=E_{\mb w}-E_{(i,i,-1)}$ per unit cell of uniform flux sectors $\mb w$ and the ground-state sector $(i,i,-1)$ as function of the coupling ratio $g$. (b) Chern numbers $\mc C$ of quasiparticle ($\varepsilon>0$) and hole ($\varepsilon<0$) bands in ground-state sector as a function of $g$. \new{The total Chern number of the quasiparticle vacuum is the sum of the Chern numbers of all negative-energy bands and only defined  for spectra with a finite quasiparticle gap. (c) Chern numbers $\mc C$ of the quasiparticle vacuum states in the uniform flux sectors $\mb w$ as a function of $g$. For $g>\sqrt{3}$ the Chern numbers vanish in all four flux sectors.}}
        \label{fig:3}
    \end{figure}

    Consistent with previous studies \cite{yao2007exact,dusuel2008perturbative,tikhonov2010quantum} \new{and a recent conjecture for arbitrary geometries~\cite{cassella2023exact}}, however, we find that the ground state belongs to the $\pi$-flux-free sector $\mb w=(i,i,-1)$ for any coupling ratio $g$, see Fig.~\ref{fig:3}(a), and that the energy gap closes for $|g|=\sqrt{3}$.
    The latter is associated with a quantum phase transition between a topological non-Abelian phase for $|g|<\sqrt{3}$ and a trivial phase \cite{yao2007exact,shi2010exotic}.
    \new{The former belongs to the same {phase} as the Kitaev honeycomb model in a weak magnetic field \cite{kitaev2006anyons} and is signaled by the finite Chern number of the ground state, which originates from the intricate topology of the energy bands shown in Fig.~\ref{fig:3}(b) and in Fig.~\ref{fig:3}(c).
    In the latter figure, we also show the Chern numbers of the fermionic vacuum states in the other uniform sectors for reference, noting that the even Chern number observed in the sector $\mb w=(i,i,1)$ is consistent with the general discussion in Ref. \cite{fuchs2020parity}.}
    
    In addition to these established results, we report the existence of perfectly flat bands of fractional fermionic excitations.
    These flat bands appear for fine-tuned values of the coupling ratio $g$ and are signaled by a vanishing bandwidth in Figs.~\ref{fig:2}(e)-(h).
    The first flat band is observed in the ground-state sector $\mb w=(i,i,-1)$ at coupling ratio and the excitation energy
    \be
        \label{eq:flat1}
        g_{1}^\star=\pm \frac{2}{\sqrt{3}},\quad\varepsilon^\star_1=\frac{4}{\sqrt{3}}|J_\trdo|,
    \ee
    respectively.
    Consequently, this flat band appears at the cusp of the fermionic gap within the topological phase, see Fig.~\ref{fig:2}(e), and is formed by compact localized matter excitations, as we demonstate in Sec.~\ref{sec:3}.

    We find additional flat bands in the high-energy flux sector $\mb w=(i,i,1)$ populated by $\pi$ flux excitations on dodecagonal plaquettes, see Figs.~\ref{fig:2}(c), (g).
    In the topological phase, the second flat band emerges at \new{the gap closing point in Fig.~\ref{fig:2}(g)}, i.e., the coupling ratio and energy 
    \be
        \label{eq:flat2}
        g_{2}^\star=\pm 1,\quad\varepsilon^\star_2=0,
    \ee
    respectively.
    Here, the energy $\varepsilon^\star_2$ is measured with respect to the fermionic vacuum of the flux sector \new{and is associated with the jump of the vacuum Chern number in Fig.~\ref{fig:3}(c)}.
    As we discuss below, this remarkable band is formed by \emph{unhybridized} compact localized MZMs with notable implications for anyonic braiding processes.

    We note that another flat band in the topologically trivial regime of the same flux sector is found for the parameters
    \be
        \label{eq:flat3}
        g_3^\star=\pm 2,\quad \varepsilon^\star_3=4\sqrt{3}|J_\trdo|.
    \ee
    As shown in Fig.~\ref{fig:2}(g), this flat band appears at the band touching point of two higher-energy bands and thus acquires an additional two-fold degeneracy.
    We rationalize this result in the following Sec.~\ref{sec:3} when deriving the conditions for the appearance of perfectly flat bands of Majorana hopping models.
    {We note that at $g=0$, additional flat bands emerge since the triangles of the star lattice are decoupled for $J_0=0$. 
    This trivial regime is not discussed further.}
    
\section{Compact localized states}
    \label{sec:3}
    The extensive degeneracy of perfectly flat bands permit linear combinations of Bloch states that form  CLS, i.e., eigenstates with support on a finite number of sites.
    It is well established that for conventional fermionic and bosonic tight-binding models, this effect stems from destructive interference of the single-particle wavefunction and typically requires fine tuning of hopping amplitudes given a specific lattice geometry \cite{sutherland1986localization,bergman2008band,chalker2010anderson}.
    This result implies that CLS do not necessitate translational invariance of the full system but that the specific conditions for their emergence are imposed on their local vicinity. 
    We may thus derive these conditions for the generic Majorana hopping model~\eqref{eq:majorana_hopping} using a real-space representation. 

    To this end, we consider linear combinations of itinerant Majoranas of the form
    \be
        \label{eq:creation_op}
        A_\Lambda^\dagger=\sum_{j\in\Lambda}\varphi_j c_j,
    \ee
    were $\Lambda$ is a finite cluster of a few lattice sites and $\varphi_j$ are complex coefficients subject to the normalization constraint $\sum_j|\varphi_j|^2={1/2}$.
    We note that operators $A_\Lambda^\dagger$ and $A_{\Lambda'}^\dagger$ corresponding to different clusters $\Lambda$ and $\Lambda'$ generally do not satisfy the anticommutation (or commutation) relations of Majorana fermions or canonical fermions (or canonical bosons) since anticommutators for overlapping clusters do not vanish \cite{bergman2008band}.

    However, we might nevertheless assume that $A_\Lambda^\dagger$ is the creation operator of a non-canonical mode such as $A_\Lambda^\dagger\ket{\emptyset}_t$, where $\ket{\emptyset}_t$ is the vacuum of quasiparticle excitations.
    Imposing that this mode is an eigenstate of $H_t$ in Eq.~\eqref{eq:majorana_hopping} with energy $\varepsilon$ relative to the quasiparticle vacuum energy yields $[H_t, A_\Lambda^\dagger]=\varepsilon A_\Lambda^\dagger$, which, in turn, translates to two distinct constraints for sites $j$ within and outside the cluster $\Lambda$.
    The latter constraint reads
    \be
        \label{eq:destructive}
        \sum_{k\in\Lambda} t_{jk}\varphi_k=0\quad\text{for all } j\notin\Lambda,
    \ee
    and reflects the requirement of destructive interference of the single-particle wavefunction $\varphi_j$ on sites coupled to the cluster.
    The second constraint imposed on sites within the cluster,
    \be
        4i\sum_{j,k\in\Lambda} t_{jk}\varphi_k c_j=\varepsilon A_\Lambda^\dagger\quad\text{for all } j\in\Lambda,
    \ee
    can be reduced to the eigenproblem 
    \be
        \label{eq:eigenproblem}
        4it_\Lambda \boldsymbol{\varphi}=\varepsilon\boldsymbol{\varphi},
    \ee
    where $\boldsymbol{\varphi}$ is a vector formed by the coefficients $\varphi_{j}$ and $t_\Lambda$ is an antisymmetric matrix involving the hopping amplitudes $t_{jk}$ for $j,k\in\Lambda$.
    
    Satisfying simultaneously both Eq.~\eqref{eq:destructive} and Eq.~\eqref{eq:eigenproblem}  generally requires a specific eigenenergy $\varepsilon$ as well as fine-tuned hopping matrix elements $t_{jk}$.
    Moreover, the connectivity of the lattice imposes additional conditions on the cluster $\Lambda$. 
    Focusing on nearest-neighbor hopping, this is because destructive interference in Eq.~\eqref{eq:destructive} requires that every neighbor of the cluster $\Lambda$ is a neighbor of at least two sites in $\Lambda$.
    On the star lattice, this excludes the boundaries of triangular plaquettes as potential clusters and singles out the dodecagonal plaquettes as the most simple candidates respecting the $\mc C_3$ rotation symmetry of the Yao-Kivelson model~\eqref{eq:model}.
    In what follows, we hence study clusters on dodecagonal plaquettes and distinguish between two physically distinct cases: plaquettes pierced by a $\pi$ flux and plaquettes without flux.

\begin{figure}
        \centering
        \includegraphics[width=\columnwidth]{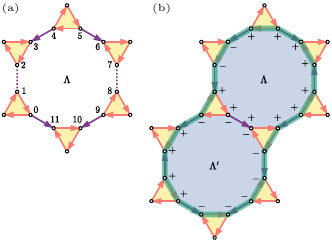}
        \caption{(a) Cluster $\Lambda$ on a dodecagonal plaquette. An arrow pointing from site $j$ to site $k$ indicates $u_{jk}=+1$. (b) Coefficients $\varphi_j=\pm$ (up to normalization) of compact localized MZMs on clusters $\Lambda$ and $\Lambda'$, see Eq.~\eqref{eq:creation_op}. The thick green loop hosts a CLS created by $(A^\dagger_\Lambda+A^\dagger_{\Lambda'})/\sqrt{2}$.}
        \label{fig:4}
    \end{figure}

    While we focus on the flux sectors $\mb w=(i,i,\pm 1)$ hosting flat bands, we note that our derivation can be easily generalized to confirm the absence of CLS in the staggered flux sectors $\mb w=(-i,i,\pm 1)$.
    As shown in Figs.~\ref{fig:2}(a), (c), the selected gauge configurations of the flux sectors of interest differ only in the values for the $z_0$-bonds.
    We therefore label sites on a plaquette using the convention in Fig.~\ref{fig:4}(a) and distinguish between the two flux sectors $\mb w=(i,i,\pm 1)$ by controlling the parameters $u_{1,2}$ and $u_{7,8}$.
    The hopping matrix $t_\Lambda$ in the resulting one-dimensional eigenproblem~\eqref{eq:eigenproblem} is detailed in Appendix~\ref{sec:b}.
    We note that while $t_\Lambda$ and $\boldsymbol\varphi$ are generally dependent on the gauge configuration, the energies of CLS and required coupling ratios $g=J_0/J_\trdo$ below are gauge-invariant.
    
\subsection{Zero flux} 
    Starting with dodecagonal plaquettes without flux, we set $u_{1,2}=1$ and $u_{7,8}=-1$, see Fig.~\ref{fig:2}(a).
    Considering the indicated signs of the hopping matrix elements, we find that the destructive interference in Eq.~\eqref{eq:destructive} readily equates the coefficients of sites on the same triangle, i.e.,
    \be
        \label{eq:intertriangle_coeff}
        \varphi_{2r+1}=\varphi_{2r}
    \ee
    where $r\in\{0,...,5\}$.
    When we impose this result on Eq.~\eqref{eq:eigenproblem}, the eigenproblem for the hopping matrix in Appendix~\ref{sec:b} is generally overdetermined.
    In fact, solutions only exist, if the systems of equations
    \be
        \label{eq:fine-tuning}
        \frac{\varepsilon}{4i}\bmat \varphi_{2r}\\\varphi_{2r+2}\emat=\bmat
            -J_\trdo&(-1)^rJ_0\\
            (-1)^{r+1}J_0&J_\trdo
        \emat
        \bmat \varphi_{2r}\\\varphi_{2r+2}\emat
    \ee
    are satisfied for every $r\in\{0,...,5\}$.
    Consequently, the energy magnitude is $|\varepsilon|=4\sqrt{J_0^2-J_\trdo^2}$ for physical solutions existing in the regime $|J_0|\geq|J_\trdo|$.
    Moreover, all coefficients merely differ by a complex phase,
    \be
        \label{eq:coeff_flux}
        \varphi_{2r+2}=(-1)^{r+1}i\mr{e}^{i\theta}\varphi_{2r},
    \ee
    where {we used the convention $\varphi_{12}=\varphi_0$ and defined}
    \be
        \label{eq:phase}
        \mr e^{i\theta}=\frac{\varepsilon+4iJ_\trdo}{4J_0}.
    \ee
    Solutions of Eq.~\eqref{eq:eigenproblem} exist if the phase accumulated by winding around the plaquette is unity, i.e., 
    \be
        \label{eq:condition_noflux}
        \mr e^{i6\theta}=1.
    \ee
    This is the condition for CLS on dodecagonal plaquettes without flux and for flat bands in the sector $\mb w=(i,i,-1)$.
    As the excitation energy $\varepsilon$ must be non-negative, we find three physical solutions.
    Besides the trivial solution $\mr e^{i\theta}=\mr{sign}(J_0)$ corresponding to the decoupled limit $J_\trdo=0$, we confirm the energy and fine-tuned coupling ratio in Eq.~\eqref{eq:flat1} for the first flat band.
    For antiferromagnetic coupling $J_0>0$, these solutions correspond to $\theta=\pm \pi/3$, while ferromagnetic coupling yields  $\theta=\pm 4\pi/3$.
    
\subsection{Finite flux}
    The above calculation can be easily adapted for dodecagonal plaquettes pierced by a $\pi$ flux.
    To this end, we note that while the sign of the bond variables $u_{jk}$ on the $z_0$-bonds alternates in Fig.~\ref{fig:2}(c), we always have $u_{1,2}=u_{7,8}$ when labeling the sites on a plaquette boundary as in Fig.~\ref{fig:4}(a).
    As destructive interference still requires coefficients of the form in Eq.~\eqref{eq:intertriangle_coeff}, one obtains solutions from above by substituting $J_0\to u_{1,2}J_0$ in Eq.~\eqref{eq:fine-tuning} for $r=0$ and $r=3$.
    While the energy magnitude is again given by $|\varepsilon|=4\sqrt{J_0^2-J_\trdo^2}$, we find that the phase difference of the coefficients is modified to
    \be
        \label{eq:coeff_flux}
        \varphi_{2r+2}=(-1)^{r+1}is_r\mr{e}^{i\theta}\varphi_{2r}, 
    \ee
    where $s_r$ is a sign defined by $s_0=u_{1,2}$, {$s_3=-u_{1,2}$} and $s_r=1$ for all other $r$, and with $\theta$ defined in Eq.~\eqref{eq:phase}.
    Equating the accumulated phase of a loop with unity yields the condition for CSL
    \be
        \mr e^{i6\theta}=-1.
    \ee
    {Notably, the change of sign with respect to Eq.~\eqref{eq:condition_noflux}, reflects the enclosed $\pi$ flux.}
    This equation has four physical solutions with non-negative energy.
    The pair of solutions $\theta=\pm\pi/2$ confirms the coupling ratio for the zero-energy flat band in Eq.~\eqref{eq:flat2}.
    By virtue of Eq.~\eqref{eq:coeff_flux}, we note that these solutions allow for purely real coefficients $\varphi_j$ which merely differ by sign, see Fig.~\ref{fig:4}(b).
    {The rescaled operator $\sqrt{2}A_\Lambda$, see Eq.~\eqref{eq:creation_op}, thus describes a Majorana \new{(zero)} mode attached to the $\pi$ flux \cite{kitaev2006anyons}.}
    \new{Going beyond uniform flux sectors, we can thus identify these CLS as the MZMs that are attached to isolated low-energy flux excitations on dodecagonal plaquettes, which we refer to as \emph{dodecagonal $\mathbbm Z_2$ vortices.}}
    Notably, given two clusters $\Lambda$ and $\tilde{\Lambda}$ on dodecagonal $\mathbbm Z_2$ vortices not adjacent to any other flux excitations, we can annihilate and create a fermionic zero mode using the canonical operators
    \be
        \label{eq:fermionic_ZM}
        {a_{\Lambda\tilde{\Lambda}}=\frac{1}{\sqrt{2}}( A_\Lambda+iA_{\tilde{\Lambda}}),\quad a_{\Lambda\tilde{\Lambda}}^\dagger=\frac{1}{\sqrt{2}}( A_\Lambda-iA_{\tilde{\Lambda}}),}
    \ee
    respectively.
    The composite objects of CLS and dodecagonal $\mathbbm Z_2$ vortex then form {compact localized Ising anyons} with analytically available wavefunctions.
    \new{We note that while MZMs attach to dodecagonal $\mathbbm Z_2$ vortices for arbitrary coupling ratios within the topological phase, they generally decay exponentially in space and are thus prone to hybridization.
    At the coupling ratio $g_2^\star=1$, however, these MZMs become CLS and thereby form perfectly delocalized fermions with exactly zero energy.}
    
    The second pair of physical solutions with $\theta=\pm \pi/6$ reproduces the results in Eq.~\eqref{eq:flat3} for the third flat band.
    As this confirms that the flat band is formed by CLS on dodecagonal plaquettes, we can rationalize the band touching point shown in Fig.~\ref{fig:2}(g).
    For this purpose, consider a finite lattice with $N$ elementary unit cells.
    While the lattice comprises $N$ dodecagonal plaquettes, a single band in the flux sector $\mb w=(i,i,1)$ hosts only $N/2$ excitations due to the aforementioned enlarged unit cell of the hopping model~\eqref{eq:majorana_hopping}.
    To populate every dodecagonal plaquette with a CLS, two flat bands thus have to coincide.
    
    On the other hand, for a low-energy excitited state with isolated dodecagonal $\mathbbm Z_2$ vortices at coupling ratio $g_3^\star=2$, two degenerate compact localized fermionic bound states will form on the vortices {with energies between the first and second quasiparticle band in Fig.~\ref{fig:2}(e)}.
    {The appearance of these bound states is consistent with the different Chern numbers of the bands, see Fig.~\ref{fig:3}(b).}

\section{Spin correlations}
    \label{sec:4}
    \begin{figure}
        \centering
        \includegraphics[width=\columnwidth]{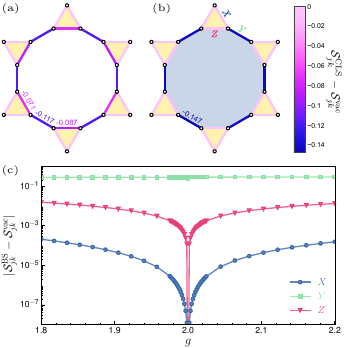}
        \caption{{Spin correlations near a dodecagonal plaquette for a system with periodic boundary conditions and linear system size $L=37$. (a) \& (b) Compact contributions $\mc{S}^{\mr{CLS}}_{jk}-\mc{S}^{\mr{vac}}_{jk}$ in Eq.~\eqref{eq:spin_CLS} indicated by the color of the corresponding bond $\braket{jk}$ for (a) a CLS in the ground-state sector $\mb w=(i,i,-1)$ for $g=g^\star_1$ and (b) a CLS for $g=g^\star_3=2$ in the excited flux sector described in the main text. Numerical values of finite contributions are annotated at the corresponding bonds. (c) Contribution $\mc{S}^{\mr{BS}}_{jk}-\mc{S}^{\mr{vac}}_{jk}$ of the bound state in the excited flux sector as a function of the coupling ratio $g$ for the spins connected by the bonds indicated in (b).}}
        \label{fig:5}
    \end{figure}
    Given the exact expressions of CLS in Sec.~\ref{sec:3}, let us investigate how the localization physics of these states manifests in physical observables.
    To this end, we compute the equal-time spin-spin correlation functions $\mc{S}^\Psi_{\alpha\beta jk}=\braket{\Psi|\sigma^\alpha_j\sigma^\beta_k|\Psi}$ for systems with no or a few flux excitations.
    As in the case of the Kitaev honeycomb model, these correlations are ultra-local for any eigenstate $\ket{\Psi}$ \cite{baskaran2007exact,knolle2014dynamics}.
    In particular, $\mc{S}^\Psi_{\alpha\beta jk}$ is only non-zero if $\alpha=\beta$ and if $j$ and $k$ are connected by a bond of type $\alpha$, which implies a one-to-one correspondence between bonds and non-zero correlations.
    The numerical value of the latter can be computed within a gauge configuration of the flux sector using \cite{baskaran2007exact}
    \be
        \label{eq:spin_correlations}
        {\mc S_{jk}^\Psi=}\mc S_{\alpha\alpha jk}^\Psi=-iu_{jk}\mc C_{jk}^{\psi},
    \ee
    where $\ket{\psi}$ is the matter sector of $\ket{\Psi}$ and an eigenstate of the resulting hopping model~\eqref{eq:majorana_hopping}, and 
    \be
        \label{eq:majorana_correlations}
        \mc C_{jk}^{\psi}=\braket{\psi|c_jc_k|\psi}
    \ee
    are (gauge-dependent) equal-time correlation functions of itinerant Majoranas.
    To determine $\mc C_{jk}^{\psi}$ for systems without translational invariance, we diagonalize the hopping model~\eqref{eq:matter_hopping} using the real-space Bogoliubov transformation detailed in Appendix~\ref{sec:a} {for a lattice with $N=L\times L$ unit cells}.
    
    We note that the local constraints on the Majorana representation~\eqref{eq:spin_majorana} imply a fermionic parity constraint which is generally sensitive to the boundary condition \cite{pedrocchi2011physical,zschocke2015physical}.
    For example, depending on the parity of physical states, a ground state of the Yao-Kivelson model~\eqref{eq:model} is given by either $\ket{\emptyset}_t$ or $a_{t,1}^\dagger\ket{\emptyset}_t$, where $\ket{\emptyset}_t$ is the {quasiparticle vacuum} for the $\pi$-flux-free sector $\mb w=(i,i,-1)$ {and $a_{t,1}^\dagger$ is the creation operator of the quasiparticle with the lowest single-particle energy, see Appendix~\ref{sec:a}}.
    In the thermodynamic limit, however, physical observables are not expected to depend on this technical detail \cite{zschocke2015physical}.

    Using this formalism, we can compute the spin correlations~\eqref{eq:spin_correlations} of CLS $A_\Lambda^\dagger\ket{\emptyset}_t$ with $A_\Lambda^\dagger$ defined in Eq.~\eqref{eq:creation_op}.
    We find that the correlations of CLS generate a \emph{compact} contribution on top of the ground-state correlations,
    \be
        \label{eq:spin_CLS}
        {\mc{S}_{jk}^{\mr{CLS}}=\mc{S}_{jk}^{\mr{vac}}-2iu_{jk}\sum_{l,m\in\Lambda}\varphi_l\varphi^*_m(\delta_{lk}\mc{C}_{mj}^{\mr{vac}}-\delta_{lj}\mc{C}_{mk}^{\mr{vac}})},
    \ee
    where $\mc{S}_{jk}^{\mr{vac}}$ and $\mc{C}_{jk}^{\mr{vac}}$ are the spin and Majorana correlations of $\ket{\psi}=\ket{\emptyset}_t$, respectively.
    This provides a characteristic signature of CLS, as shown in Fig.~\ref{fig:5}.
    We note, however, that for a system tuned to the flat band in the ground-state sector, see Fig.~\ref{fig:5}(a), the spin correlations of fermionic excitations will not be compact but extend over the whole system.
    This is because CLS provide a non-orthogonal basis of the extensively degenerate flat-band manifold, while any orthogonal basis formed by canonical fermions will not feature compact localization.

    The situation is different for a system with a few isolated flux excitations trapping CLS on dodecagonal $\mathbbm Z_2$ vortices  at the coupling ratios $g_2^\star$ and $g_3^\star$ in Eqs.~\eqref{eq:flat2} and \eqref{eq:flat3}, respectively.
    Starting with the latter coupling ratio in the trivial phase, these CLS are orthogonal canonical fermions in the inter-band gap, see Sec.~\ref{sec:3}.
    As a result, these bound state excitations cannot mix with extended states or bound states on non-adjacent flux excitations and permit the observation of the compact spin-spin correlations~\eqref{eq:spin_CLS}. 
    We illustrate this result in Fig.~\ref{fig:5}(b) for the vortex sector obtained from the ground-state sector $\mb w=(i,i,-1)$ by inserting flux excitations in a single dodecagonal plaquette and a well-separated up-pointing triangular plaquette using a gauge configuration that is consistent with the one assumed in Section~\ref{sec:3} and Fig.~\ref{fig:4}(b).
    Notably, we find that the compact contributions~\eqref{eq:spin_CLS} are only finite for inter-triangle bonds on the vortex boundary and vanish for any spin pair on any triangle.
    This result stems from the destructive interference necessary for the formation of the CLS and represents a distinct difference to the behavior of generic bound states that are exponentially localized near flux excitations on a length scale controlled by the bound-state energy.

    To confirm this notion, we use the expressions detailed in Appendix~\ref{sec:a} to compute the spin-spin correlations~\eqref{eq:spin_correlations} of the single-particle bound-state excitation $\ket{\psi}=a_{t,\nu}^\dagger\ket{\emptyset}_t$ for the described vortex sector.
    Since the bound states near the flux excitations in the triangular and dodecagonal plaquettes have different energies for $g\neq g_2$, they do not hybridize and can be readily distinguished.
    Fig.~\ref{fig:5}(c) shows the resulting correlations $\mc S^{\mr{BS}}_{jk}$ of selected spin pairs for the bound state near the dodecagonal $\mathbbm Z_2$ vortex as a function of the coupling ratio $g$.
    We find that the intra-triangle correlations drop super-exponentially upon tuning $g$ to the coupling ratio $g_3^\star$ and thus provide a clear-cut benchmark of compact localization.

    Finally, we may discuss dodecagonal $\mathbbm Z_2$ vortices for the coupling ratio $g_2^\star$ in the topological phase.
    In this case, two separated CSL form ideal fermionic zero modes $\ket{\psi_{\mr{MZM}}}=a^\dagger_{\Lambda\tilde{\Lambda}}\ket{\emptyset}_t$ created by the operators $a_{\Lambda\tilde{\Lambda}}^\dagger$ in Eq.~\eqref{eq:fermionic_ZM}.
    {According to the general understanding of topological order} \cite{kitaev2006anyons}, such states cannot be distinguished from the vacuum $\ket{\emptyset}_t$ by any local operators.
    Using the exact solutions in Sec.~\ref{sec:3}, one can easily confirm this fundamental prediction for the spin-spin correlations, $\mc S^{\mr{MZM}}_{jk}=\mc S^{\mr{vac}}_{jk}$, thereby demonstrating the complete absence of hybridization of MZMs.

\section{Discussion}
    \label{sec:5}
    This work establishes the existence of perfectly flat quasiparticle bands populated by compact localized states in the Yao-Kivelson model.
    The formation of CLS stems from destructive quantum interference on the triangles decorating an underlying honeycomb lattice.
    In this sense, they share their origin with the spontaneous breaking of time-reversal symmetry, which is induced by the existence of Wilson loops with odd length \cite{yao2007exact,kitaev2006anyons}.
    It would be interesting to explore the relation of flat bands and odd-length Wilson loops on more general grounds or for different lattice geometries and models.
    To this end, it should be possible to adapt our formalism based on the generic Majorana hopping model~\eqref{eq:majorana_hopping}.
    \new{While the Yao–Kivelson model is a theoretically motivated construction with an exact chiral spin-liquid ground state, this procedure may also enable the study of flat bands in related models,
    thereby broadening the possibilities for realizing them in real materials
    \cite{trebst2022kitaev}.}
    We note that the construction of CLS becomes straightforward if the star lattice is further decorated with triangles.
    In this case, the conditions for CLS on plaquettes with even length $\ell=2M$ generalize to $\mr{e}^{iM\theta}=\pm 1$, where $\theta$ is the phase defined in Eq.~\eqref{eq:phase} and the sign $\pm$ is determined by the enclosed flux. 
    
    On the other hand, we might attempt to construct CLS on larger loop-like clusters of the star lattice.
    In the context of conventional tight-binding models with conserved particle number, the existence of CLS on these larger loops is associated with a topologically protected touching point with a dispersive band \cite{bergman2008band}.
    An important consequence of this band touching point is an algebraic decay of the projector onto the flat band in real space, which entails critical states in the presence of weak disorder \cite{chalker2010anderson}.
    In our case, however,  the construction of larger clusters is more subtle, since all flat bands of the Yao-Kivelson model are fully gapped from dispersive bands.
    Physically, this subtlety can also be attributed to the fact that any larger loop-like cluster should enclose the same amount of flux (modulo $2\pi$), taking into account the triangular plaquettes pierced by the flux $\pm\pi/2$.
    In the case of the zero-energy flat band in Eq.~\eqref{eq:flat3}, this constraint is satisfied for the construction exemplified in Fig.~\ref{fig:4}(b).
    There, the sum $(A^\dagger_\Lambda+A^\dagger_{\Lambda'})/\sqrt{2}$, with $A_\Lambda^\dagger$ in Eq.~\eqref{eq:creation_op}, {describes a compact localized MZM} on the larger loop-like cluster $(\Lambda \cup \Lambda') \setminus (\Lambda \cap \Lambda')$ enclosing the flux $3 \pi\text{ mod }2=\pi$.
    We leave a more systematic analysis of other loop-like clusters to future studies {but note that CLS on loops winding around a system with periodic boundary conditions are not expected, see Ref.~\cite{bergman2008band}}.
    \new{As a consequence, we anticipate that finite bond disorder does not induce critical wavefunctions~\cite{chalker2010anderson} but exponentially localized eigenstates with a comparably sharp bandwidth.}

    Another interesting facet of flat bands is their susceptibility to interaction-driven instabilities that can induce phase transitions to unconventional states of matter upon tuning the filling to the flat-band energy.
    In the case of the Yao-Kivelson model~\eqref{eq:model}, this may be achieved by reorganizing the energetic hierarchy among the flux sectors.
    The most obvious approach to this end is to couple the Hamiltonian to the plaquette operators $W_p$ in Eq.~\eqref{eq:wilson} using \cite{tikhonov2010quantum}
    \be
        \label{eq:coupling_flux}
        H_\mr{flux}=J_{12}\sum_{p_0}W_{p_0}+J_6\sum_{(p_\trdo,q_\trup)}W_{p_\trdo}W_{q_\trup},
    \ee
    where the first sum runs over dodecagonal plaquettes $p_0$ and the second sum runs over pairs of down-pointing triangles $p_\trdo$ and neighboring up-pointing triangles $q_\trup$.
    For appropriate values for the coupling constants $J_{12}<0$ and $J_6>0$, the ground state resides in the flux sector $\mb w=(i,i,+1)$.
    Since the band structure of fermionic excitations is not affected by $H_\mr{flux}$ and the conditions for flat bands remain unchanged, we obtain a \emph{quantum} spin liquid with extensive ground-state degeneracy at the magic coupling ratio $g^\star_2=1$.
    This exotic state should be highly unstable, and we predict a phase transition in the presence of interactions{, see Ref.~\cite{yogendra2025fractional}}.
    While the term in Eq.~\eqref{eq:coupling_flux} seems artificial, we note that this scenario might also emerge in the presence of more natural interactions.
    This is plausibly inferred from works on the Kitaev honeycomb model showing that four-spin interactions can induce flux crystallization, i.e., ground states in sectors with nontrivial flux patterns \cite{zhang2019vison,fuchs2020parity,chulliparambil2021flux,alspaugh2024effective}. 
    In the case of the Yao-Kivelson model, if a flux pattern includes isolated dodecagonal $\mathbbm Z_2$ vortices, it hosts a zero-energy flat band at $g=g^\star_2$ that is prone to instabilities.
    \new{We note that no instability is expected in a system hosting the finite-energy flat band found in the sector $\mb w=(i,i,-1)$. 
    Instead, the system should remain in the same phase under sufficiently small perturbations to the spin model~\eqref{eq:model} such as the additional exchange terms relevant in material realizations \cite{winter2017models, hermanns2018physics,trebst2022kitaev}.
    While such perturbations generally break integrability and obstruct the formation of CSL, our results could offer valuable insight into the behaviour of the bandwidth of fractional excitations within the spin-liquid phase.}
    
    CLS do not only emerge in extensively degenerate flat bands of systems with translational invariance, but also when their respective conditions are met locally.
    In the present work, the most interesting example of this scenario is given by compact localized MZMs on isolated dodecagonal $\mathbbm Z_2$ vortex excitations at the magic coupling ratio $g_2^\star$.
    The composite object of MZM and $\mathbbm Z_2$ vortex corresponds to an Ising anyon with non-Abelian exchange statistics.
    While braiding via particle exchange typically requires a sufficiently large separation of anyons to suppress hybridization \cite{kitaev2006anyons,nayak2008non}, we find that hybridization is completely absent for compact localized Ising anyons.
    In particular, if they are separated by a single dodecagonal plaquette, they can be described by the orthogonal canonical zero modes in Eq.~\eqref{eq:fermionic_ZM}.
    We infer that non-Abelian braiding of Ising anyons with minimum distance is possible and can therefore be realized in quantum simulations with currently accessible system sizes \cite{google2023non,xu2023digital,xu2024non,iqbal2024non,iqbal2025qutrit,song2025shortcuts}.
    Existing proposals for the control of anyons in the Kitaev honeycomb model are based on adiabatic modulation of the energy coupling \cite{bolukbasi2012rigorous}, local time-dependent magnetic fields \cite{harada2024real} or local electric probes \cite{aasen2020electrical,pereira2020electrical,bauer2023scanning,klocke2024spin}.
    We note that in our case, any braiding protocol requires careful analysis on how and to what degree compact localization is obstructed. 
    Similar considerations are necessary when considering detuning from the magic coupling ratio.
    We leave this problem to future studies, but we are confident that our work can guide the quantum simulation of non-Abelian braiding processes and lays the groundwork for the exploration of flat-band phenomenology in chiral spin liquids.
    
\begin{acknowledgments} 
    {We acknowledge early results by Matts Nissen and Vatsal Dwivedi that inspired this work.}
    We thank Rodrigo Pereira, Han Yan and Yukitoshi Motome for helpful discussions.
    J.R. acknowledges support from the Deutsche Forschungsgemeinschaft (DFG, German Research Foundation), within Project-ID 277101999 CRC 183 (Project A04).
    T.B. acknowledges support from the Japan Society for the Promotion of Science under a JSPS Postdoctoral Fellowship for Research in Japan (Short-Term). 
    All data underlying the figures in this work are available at Zenodo~\cite{Zenodo}.
    
\end{acknowledgments}
\appendix
\section{Solution of free-fermion problem}
    \label{sec:a}
    Here, we outline the diagonalization of the hopping model~\eqref{eq:matter_hopping} and the computation of the Majorana-Majorana correlation functions~\eqref{eq:majorana_correlations} introduced in Section~\ref{sec:4}.
    Assuming a system on a lattice with $N=L\times L$ unit cells, we can recast Eq.~\eqref{eq:matter_hopping} to
    \be
        \label{eq:matrix_hopping}
        H_t=\frac{1}{2}\bmat f\\f^\dagger\emat^\dagger \bmat
            A_t&B_t\\
            -B_t^*&-A_t^*
        \emat\bmat f\\f^\dagger\emat,
    \ee
    where $f$ ($f^\dagger$) is a column vector composed of all annihilation (creation) operators $f_m$ ($f_m^\dagger$) and $A_t$ and $B_t$ are $3N\times 3N$ matrices with matrix elements (dropping the label $t$ in what follows)
    \be
        A_{mn}=\delta_{mn}\mu_m+2i\mr{Im}\lb\tau^+_{mn}\rb,\quad B_{mn}=2\tau^-_{mn}.
    \ee
    The parameters $\tau^\pm_{mn}$ and $\mu_m$ are defined below Eq.~\eqref{eq:matter_hopping}.
    The key step is a unitary transformation $W$ to a new species of fermions,
    \be
        \bmat f\\f^\dagger\emat=W\bmat a\\a^\dagger\emat, 
    \ee
    where $a$ ($a^\dagger$) is a column vector composed of all annihilation (creation) operators $a_\nu$ ($a_\nu^\dagger$).
    The canonical anticommutation relations of fermions imply that the transformation is of the form \cite{blaizot1986quantum}
    \be
        \label{eq:bogoliubov}
        W=\bmat
            X^*&Y^*\\
            Y&X
        \emat,
    \ee
    where $X$ and $Y$ are $3N\times 3N$ matrices.
    Given the particle-hole symmetry of the hopping model~\eqref{eq:matter_hopping}, we can choose $W$ such that
    \be
        W^\dagger \bmat
            A&B\\
            -B^*&-A^*
        \emat W=\mr{diag}\lb \varepsilon_1, ..., \varepsilon_{3N},-\varepsilon_1, ..., -\varepsilon_{3N}\rb,
    \ee
    where the single-particle energies $\varepsilon_\nu$ are ordered according to $0\leq\varepsilon_{1}\leq ... \leq \varepsilon_{3N}$.
    {Inserting the Bogoliubov transformation~\eqref{eq:bogoliubov} in Eq.~\eqref{eq:matrix_hopping} yields the diagonalized Hamiltonian (reintroducing the label $t$)
    \be
        \label{eq:diagonal_hopping}
        H_t=\sum_{\nu=1}^{3N} \varepsilon_{t\nu} \lb a_{t\nu}^\dagger a_{t\nu}-\frac{1}{2}\rb,
    \ee
    where the vacuum energy corresponds to the energy of the flux sector,
    \be
        E_t=-\frac{1}{2}\sum_{\nu=1}^{3N}\varepsilon_{t\nu}.
    \ee}
    Given this result, we can compute the Majorana-Majorana correlations~\eqref{eq:majorana_correlations} of energy eigenstates.
    To this end, we represent itinerant Majoranas in terms of the matter fermions in Eq.~\eqref{eq:matter_pairing} using $c_{m,\trdo}=f_m+f_m^\dagger$ and $c_{m,\trup}=-i\lb f_m-f_m^\dagger\rb$, and perform the Bogoliubov transformation~\eqref{eq:bogoliubov}.
    Evaluating expectation values with respect to the vacuum $\ket{\emptyset}$ annihilated by the quasiparticle operators $a_\nu$ yields the correlations
    \begin{subequations}
        \begin{align}
            \label{eq:corr_vacuum}\mc{C}_{m,\trdo,n,\trdo}^{\mr{vac}}&=\delta_{mn}+2i\imag\lb YY^\dagger+YX^T\rb_{mn}\\
            \mc{C}_{m,\trup,n,\trup}^{\mr{vac}}&=\delta_{mn}+2i\imag\lb YY^\dagger-YX^T\rb_{mn}\\
            \mc{C}_{m,\trdo,n,\trup}^{\mr{vac}}&=i\delta_{mn}-2i\real\lb YY^\dagger-YX^T\rb_{mn},
        \end{align}
    \end{subequations}
    where $(...)_{mn}$ specifies the matrix elements of the matrix in the brackets.
    From this expression we can compute the gauge-invariant spin-spin correlations~\eqref{eq:spin_correlations} of the ground state and of the CLS in Eq.~\eqref{eq:spin_CLS}.
    We note that the latter also involve correlation functions $\mc C_{jk}$ of sites $j$ and $k$ that are not nearest neighbors.
    
    To compare CLS to generic fermionic bound states, we may also evaluate the correlations $\mc C^\nu_{jk}$ of single-particle excitations of the form $a_{\nu}^\dagger\ket{\emptyset}$.
    The resulting contributions $\delta\mc{C}_{jk}^{\nu}=\mc{C}_{jk}^{\nu}-\mc{C}_{jk}^{\mr{vac}}$ on top of the vacuum correlations in Eq.~\eqref{eq:corr_vacuum} are given by
    \begin{subequations}
        \begin{align}
            \delta\mc{C}_{m,\trdo,n,\trdo}^{\nu}&=2i\imag{\lsb \lb X_{m \nu}+Y_{m \nu}^*\rb\lb X_{n \nu}^*+Y_{n \nu}\rb\rsb}\\
            \delta\mc{C}_{m,\trup,n,\trup}^{\nu}&=2i\imag{\lsb \lb X_{m \nu}-Y_{m \nu}^*\rb\lb X_{n \nu}^*-Y_{n \nu}\rb\rsb}\\
            \delta\mc{C}_{m,\trdo,n,\trup}^{\nu}&=-2i\real{\lsb \lb X_{m \nu}+Y_{m \nu}^*\rb\lb X_{n \nu}^*-Y_{n \nu}\rb\rsb}.
        \end{align}
    \end{subequations}
    For results shown in Fig.~\ref{fig:5}(c), we choose the label $\nu$ to describe the bound state within the inter-band gap and near the dodecagonal $\mathbbm Z_2$ vortex (using the modified notation $\mc{C}_{jk}^{\mathrm{BS}}=\mc{C}_{jk}^{\nu}$).
    
\section{Hopping on dodecagonal plaquette}
    \label{sec:b}
    Here, we specify the hopping matrix $t_\Lambda$ used in Section~\ref{sec:3}.
    This matrix comprises the elements $t_{jk}$ describing hopping on a dodecagonal plaquette for the gauge configurations indicated in Fig.~\ref{fig:4}(a).
    As specified in the main text, we have $t_{jk}=-u_{jk}J_\chi=-t_{kj}$ (with $J_\trdo=J_\trup$) for a bond of type $\alpha_\chi$ with $\alpha\in\{x,y,z\}$ and $\chi\in\{\trdo,\trup,0\}$.
    Using the convention for the labels $j$ in Fig.~\ref{fig:4}(a), we obtain
    \begin{widetext} 
        \be
        t_\Lambda = \left(\begin{array}{*{12}{c}}
            0  & J_\trdo  & 0      & 0      & 0      & 0      & 0      & 0      & 0      & 0       & 0       & -J_0      \\
            -J_\trdo  & 0  & -u_{1,2}J_0  & 0      & 0      & 0      & 0      & 0      & 0      & 0       & 0       & 0      \\
            0      & u_{1,2}J_0  & 0  & J_\trdo  & 0      & 0      & 0      & 0      & 0      & 0       & 0       & 0      \\
            0      & 0      & -J_\trdo  & 0  & J_0  & 0      & 0      & 0      & 0      & 0       & 0       & 0      \\
            0      & 0      & 0      & -J_0  & 0  & J_\trdo & 0      & 0      & 0      & 0       & 0       & 0      \\
            0      & 0      & 0      & 0      & -J_\trdo  & 0  & -J_0  & 0      & 0      & 0       & 0       & 0      \\
            0      & 0      & 0      & 0      & 0      & J_0  & 0  & J_\trdo  & 0      & 0       & 0       & 0      \\
            0      & 0      & 0      & 0      & 0      & 0      & -J_\trdo  & 0  & -u_{7,8}J_0  & 0       & 0       & 0      \\
            0      & 0      & 0      & 0      & 0      & 0      & 0      & u_{7,8}J_0  & 0  & J_\trdo   & 0       & 0      \\
            0      & 0      & 0      & 0      & 0      & 0      & 0      & 0      & -J_\trdo & 0 & -J_0  & 0      \\
            0      & 0      & 0      & 0      & 0      & 0      & 0      & 0      & 0      & J_0  & 0  & J_\trdo \\
            J_0      & 0      & 0      & 0      & 0      & 0      & 0      & 0      & 0      & 0       & -J_\trdo  & 0
        \end{array}\right),
        \ee
    \end{widetext}
    where the bond variables $u_{1,2}=-u_{2,1}$ and $u_{7,8}=-u_{8,7}$ control the flux in the enclosed plaquette by virtue of the eigenvalue of the corresponding plaquette operator~\eqref{eq:wilson} $w=u_{1,2}u_{7,8}$.
\bibliography{references}
\end{document}